# Ultrafast pulse generation in a mode-locked Erbium chip waveguide laser

CHAMPAK KHURMI[1,4], NICOLAS BOURBEAU HÉBERT[2], WEN QI ZHANG[1], SHAHRAAM AFSHAR V.[1], GEORGE CHEN[1], JÉRÔME GENEST[2], TANYA M. MONRO[1,3], DAVID G LANCASTER[1,3,5]

[1]*Laser Physics and Photonic Devices Laboratory, University of South Australia, SA, Australia*

[2]*Centre d'optique, photonique et laser, Université Laval, Québec G1V 0A6, Canada*

[3]*Red Chip Photonics, Pty Ltd, Adelaide, SA, Australia*

[4]*champak.khurmi@gmail.com*

[5]*david.lancaster@unisa.edu.au*

**Abstract:** We report mode-locked ~1550 nm output of transform-limited ~180 fs pulses from a large mode-area (diameter ~ 50 μm) guided-wave erbium fluorozirconate glass laser. The passively mode-locked oscillator generates pulses with 25 nm bandwidth at 156 MHz repetition rate and peak-power of 260 W. Scalability to higher repetition rate is demonstrated by transform-limited 410 fs pulse output at 1.3 GHz. To understand the origins of the broad spectral output, the laser cavity is simulated by using a numerical solution to the Ginzburg-Landau equation. This paper reports the widest bandwidth and shortest pulses achieved from an ultra-fast laser inscribed waveguide laser.

## 1. Introduction

Over the past few decades, sources of ultrafast laser pulses have progressed from complex and specialized systems to compact and robust tabletop instruments [1]. Because of the high peak-powers and broad spectrum, ultrafast lasers are applicable to emerging applications such as non-linear microscopy [2-4], telecommunication [5, 6], frequency combs [7, 8] and high-speed digitization [8]. To enable instruments powered by ultra-fast lasers to be suitable for use outside the laboratory, new designs for portable and robust ultrafast lasers are required. An important but underrepresented regime is ultra-short pulses with high repetition rates (0.1 to 2 GHz) in various spectral domains with high-peak powers along with compact and reliable architectures [9-12].

We have previously reported a new compact waveguide laser architecture, where ultra-fast laser (ULI) inscribed waveguides are written into chips of rare-earth doped fluorozirconate glass. By tailoring the rare-earth dopant and waveguide design we have achieved efficient laser operation throughout the near to short-wave infrared spectral domain, covering 1.1, 1.5, 1.9, 2.1 and 2.9 μm [13-17], and demonstrated extended laser tuning ranges of up to 253 nm for thulium doped ZBLAN. By combining this wavelength flexibility, broad gain bandwidths, and the intrinsic large mode-area of these depressed-cladding guiding structures, the essential requirements are met to realise a compact laser geometry that is ideal for mode-locked applications and cover specific bands across the 1-3 μm spectral domain. Here, we present a mode-locked laser that demonstrates proof of principle for this wavelength flexible architecture. Specifically, we have achieved passive ultrafast mode-locked operation based on Erbium doped Fluorozirconate glass [15, 18] with ULI large mode-area waveguides (50 μm diameter) [13].

Erbium has been used as the preferred dopant in different host materials to generate ultrafast mode-locked pulses ranging from several picoseconds to ~100 fs in the 1.5 μm wavelength domain [11, 19-25]. Generally, in the 1.5 μm band, the existing mode-locked lasers available to cover the 100 MHz to multi-GHz repetition rate are solid-state laser designs with short laser cavities (L ~ < 0.5 m) [26]. However, the requirement for robust single transverse mode operation of solid-state lasers to ensure mode-locking is challenging due to high sensitivity to cavity and mirror alignment, power dependent thermal lensing, and the precise overlap of the pumped gain region and the resonator mode. The guided-wave confinement of fiber like structures in the bulk gain media solves these alignment issues of unguided solid-state lasers. While 'all fiber' lasers are desirable for their compact and reliable design, their high gain, low Q-factor, and high-nonlinearity due to tight light confinement (mode area ~ $0.5 \times 10^{-6}$ cm$^2$) over an extended distance leads to enhanced non-linear effects such as self-phase modulation (SPM) [27] and require dispersion compensation [20].

Recently, we reported a widely tunable (~100 nm) Er-Yb-Ce doped fluoro-zirconate waveguide laser [14] with relatively large mode-area (~ $11 \times 10^{-6}$ cm$^2$). The wide and flat gain profile of Er$^{3+}$ in ZBLAN glass [14, 28, 29] combined with large mode-area waveguides make them promising candidates for peak-power scalable ultrafast laser devices.

In this paper, we demonstrate a passively mode-locked Er-Yb-Ce: ZBLAN chip laser oscillator. This waveguide-based laser generates sub 200 fs transform-limited pulses at 156 MHz repetition rate with a 25 nm spectral bandwidth. This is the widest bandwidth and shortest pulses achieved in an ultra-fast laser inscribed waveguide laser.

Numerical simulations of this mode-locked laser operating in the normal dispersion regime (ZBLAN has a zero dispersion λ of 1.6 µm [18]) suggest that the large mode-locked bandwidth results from the efficient use of the gain spectrum of erbium [14, 28, 29], instead of spectral broadening from nonlinear effects as reported earlier in Er$^{3+}$ doped fibre lasers [30]. This is conceptually significant since this design takes the advantages of guided laser geometry without the nonlinear effects.

## 2. Experiment

The glass substrate is Er$^{3+}$ and Yb$^{3+}$ co-doped (0.5 mol. % Er$^{3+}$ and 2 mol. % Yb$^{3+}$) ZBLAN glass. Cerium (5 mol. %) is added to increase the branching ratio from the Er$^{3+}$ $^4$I$_{11/2}$ pump level to the laser upper state $^4$I$_{13/2}$ [7]. Fluorozirconate glasses such as ZBLAN have high transparency from the UV to ~ 4 µm [31]. Combined with high rare-earth solubility and low phonon energies they make excellent candidates for compact and efficient waveguide lasers. Depressed cladding waveguides are directly inscribed in the glass substrate (length = 13 mm) by ultrafast laser pulse inscription (5 MHz, < 250 fs, λ = 520 nm), produced from a frequency-doubled laser (IMRA, DE0210) [8]. Further details about the ZBLAN waveguides and ultrafast laser inscription methods can be found in previously published work [32-34].

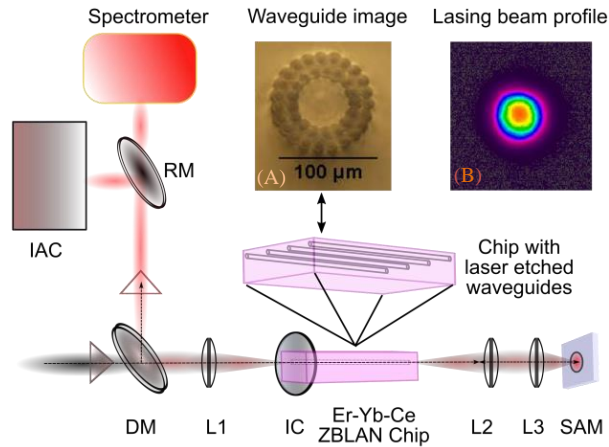

Fig. 1. Schematic of the passively mode-locked ultrafast Er-Yb-Ce-ZBLAN waveguide laser. DM: Dichroic Mirror, L$_1$: F = 30 mm, IC: Input coupler (R$_{1550}$ = 95%); L$_2$: F = 25 mm; L$_3$: (F = 4.5 mm); SAM: Semiconductor based Saturable Absorber; RM: retractable mirror to get intensity autocorrelation traces; optical spectrum and rf spectrum. Inset (A) image of the laser etched waveguide with double-ring structure (waveguide diameter ~ 50 um) and (B) nearfield beam-profile of the mode-locked laser output.

The experimental setup is shown in Fig. 1. A fiber-coupled diode laser (Thorlabs, model # BL976-PAG900) is used as the pump. The pump beam is focussed into the ZBLAN waveguide by an achromatic lens ($L_1$, Focal length (F) = 30mm) through an input/output coupler (IC, R=95%). To set the repetition rate, the laser oscillator cavity length could be easily varied between ~0.1 to 1.0 m by using two lenses in a relay configuration: $L_2$ (AR coated achromatic lens with F = 30 mm) and lens $L_3$ (F = 4.5 mm) to re-image the mode onto the SAM (BATOP Gmbh, SAM-1550-15-12ps-4). The mode-locked output is reflected out of the beam path by a dichroic mirror (DM) and passes through a combination of half-wave plate (HWPL) and polarization beam-splitter (PBSL) to get the autocorrelation trace and spectrum.

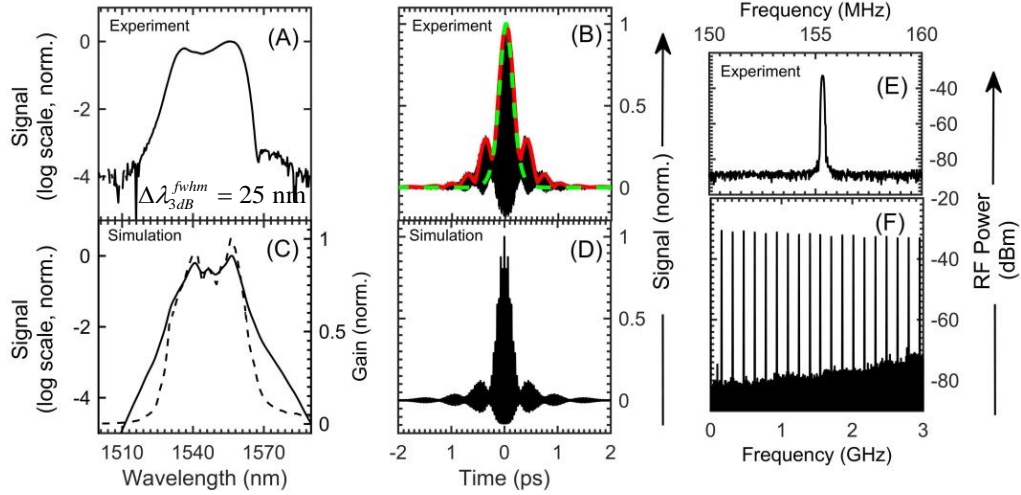

Fig. 2. Mode-locked output of the Er-Yb-ZBLAN waveguide oscillator at 156 MHz repetition rate. Experimentally observed mode-locked (A) spectrum with $\Delta\lambda_{fwhm\,(3dB)}$ = 25 nm bandwidth and (B) auto-correlation trace (black, solid line), Fourier transform of the mode-locked spectrum (red, dashed line) and $f(t)$ fit (green, dotted line) to the autocorrelation trace indicates $\tau_p^{\exp}$ = 180 fs. Exact match between the experimentally observed auto-correlation and the Fourier transform of the observed mode-locked spectrum suggests transform-limited pulse generation from the waveguide oscillator. (C) Simulated mode-locked spectrum (solid line, y-axis on left side) and simulated gain profile (dashed line, y-axis on right side) and (D) simulated auto-correlation trace from numerical model (refer eqn. 1). (E) shows the zoomed-in RF spectrum indicating single-pulse operation (F) Full scale (up to 3 GHz) RF spectrum of mode-locked pulses (RBW = 100 kHz).

Fig. 2 (A-B) shows the experimentally observed optical spectrum and auto-correlation trace of the mode-locked output at 156 MHz (cavity length ~ 96 cm), respectively. Up to ~ 5 mW of power was achieved when the laser was mode-locked. The spectrum of the mode-locked pulses is recorded by using an optical spectrum analyser (MS9470A, Anritsu Corp.). The experimental auto-correlation trace Fig. 2 (B) (black - solid line), of the mode-locked pulses was obtained by using APE Pulse Check auto-correlator (A.P.E. Angewandte Physik & Elektronic GmbH). The FWHM measured from Fig. 2 (A) is $\Delta\lambda_{fwhm}$ = 25 nm. Fig. 2 (B) (red-solid line), shows the Fourier transform of the mode-locked phase-less power spectral density. It is evident from the close fit that the waveguide laser oscillator generates transform-limited pulses. As $\text{sech}^2$ pulse shapes are extensively used to describe mode-locked pulses, we used a typical fit function [35] $f(t) = \left[\text{sech}^4\left(t/2.4445T\right)/3T\right]$, where T is the pulsewidth when $f(t)$ is decreased by $1/e^2$, to measure the pulse-width from the experimentally observed autocorrelation trace as shown in Fig. 2 (B). The dashed-line (green) in Fig. 2 (B) shows the $f(t)$ fit to the envelope of the experimental data (solid, black). From the fit, the autocorrelation

width ($\tau_{AC}$) was estimated to be 277 fs by using $\tau_{AC}= 2.720T$. It results in a pulse width of $\tau_p^{exp} = 180$ fs, such that $\tau_p^{exp} = (\tau_p^{AC}/1.54)$.

We now consider the origin of the broad mode-locked spectrum. The RF spectrum of the mode-locked pulses is shown in Fig. 2 (E-F). This data supports stable mode-locking operation, and negates the presence of multiple-pulses (which could appear as a broader bandwidth). The observed spectrum is not enhanced by nonlinear processes either since the waveguide mode-area is approximately $\sim 20 \times 10^{-6}$ cm$^2$, which leads to an estimated nonlinear coefficient of $\gamma = (2\pi n_2/\lambda A_{eff}) \approx 3 \times 10^{-5}$ W$^{-1}$m$^{-1}$ at $\lambda = 1.55$ µm for a nonlinear ZBLAN refractive index of $1.5 \times 10^{-20}$ m$^2$/W. For a waveguide length of 13 mm and ~2 kW intra-cavity peak intensity a negligible roundtrip nonlinear phase change of $3 \times 10^{-4}$ $\pi$ radians is estimated. The exact match between the Fourier transform of the mode-locked spectrum and the experimentally observed autocorrelation trace (Fig. 2 (B)) also supports this insignificant phase contribution. The dispersion of the ZBLAN waveguide is estimated to be low at ~ 7.4 ps$^2$/mm at 1550 nm by using the Sellmeier equation [31]. The waveguide dispersion is assumed close to the material dispersion due to the large mode area. Therefore, we conclude that multiple-pulse operation, non-linear effects such as SPM and higher order dispersion effects do not have significant contributions to generate broad spectral output.

## 3. Numerical model based on Ginzburg-Landau Equation

To confirm the origins of this broadband spectral profile, we used a numerical model based on the Ginzburg-Landau equation. The results from numerical simulations are shown in Fig. 2 (C) and (D) and represent the simulated mode-locked spectrum, gain profile and the auto-correlation trace respectively. Based on the numerical solution to the Ginzburg-Landau equation, which qualitatively agrees with experimental results, the wide and flat gain profile of Er$^{3+}$ (dashed line, Fig. 2 (C)) in ZBLAN glass [14, 29] accounts for the observed broad spectral output of the mode-locked pulses.

$$\frac{\partial u}{\partial z} = i\frac{\beta_2}{2}\frac{\partial^2 u}{\partial t^2} + i\gamma |u|^2 u + \frac{g(u)}{2}u$$

$$g(u) = \frac{g_0(\omega)}{1+\frac{\int |u|^2 dt}{E_{sat}}}$$

$$g_0(\omega) = N_2 \sigma_e(\omega) - N_1 \sigma_a(\omega) \quad (1)$$

$$u'(t) = \eta \left[1 - \frac{q_0}{1+\frac{|u(t)|^2}{P_{sat}}}\right] u(t)$$

In the model, we use a split-step Fourier method to solve the Ginzburg-Landau equation [36], Eqs. (1), for the pulse propagation through the cavity. In these equations, $u$ is the envelope of the laser electric field, $g(u)$ is the saturable gain, z is propagation distance, $\gamma$ is the nonlinear coefficient, $t$ is time, $\beta_2$ is the second-order dispersion, $E_{sat}$ is the saturation energy of the gain medium, $N_1$ is ground state population, $N_2$ is excited state population, $\sigma_e$ and $\sigma_a$ are the emission and absorption cross-sections respectively [14, 29], $u'$ is the pulse envelope after the saturable absorber (SA), $q_0$ is the modulation depth of the SA, $P_{sat}$ is the saturation power of SA, and η is the square root of the non-saturate reflectance. The effect of the change in gain function $g(u)$ with respect to the pump power is represented by $N_2/N$ (such that $N = N_1 + N_2$), as this ratio is proportional to the pump power.

In the simulation, the process starts from Gaussian white noise. The signal passes through the gain medium first then the phase of the signal is modified based on the dispersion properties of the lenses (lenses $L_2$ and $L_3$ in Fig. 1). A saturable absorber operation is then applied and followed by the phase modification due to the dispersion of $L_2$ and $L_3$ and propagation through the gain medium again to mimic the linear cavity. After one round trip, a multiplier of 0.95 is applied to simulate the 5% out-coupling loss. The rest of the cavity is assumed to be lossless. The same procedure repeats until the spectral amplitude of the pulse stabilizes within 0.0001%.

In this work, a range of $N_2/N$ ratios were simulated. A good qualitative match between theoretical simulations and experimental data is observed when $N_2/N=0.64$. The similarity between the simulated pulse spectrum and erbium ZBLAN gain profile to the experimental observation (refer Fig. 2 (A) and (C)) with characteristic shoulders at ~1530 and 1550 nm indicates that the mode-locked output is governed by the erbium ZBLAN gain profile due to the low dispersion and non-linearity in the cavity.

By shortening this mode-locked cavity (with trivial re-alignment) we have also achieved stable mode-locked operation at 1.3 GHz, achieving transform-limited pulsewidths of $\tau_p^{\text{exp}} = 410\,\text{fs}$ as shown in Fig. 3 (A-B). Fig. 3 (B) shows the mode-locked spectral power density. Fig. 3 (A) shows experimentally observed autocorrelation trace (black, solid line), Fourier transform (red, triangle) of the phase-less spectral power density and $f(t)$ fit (blue, diamond) to the experimental autocorrelation.

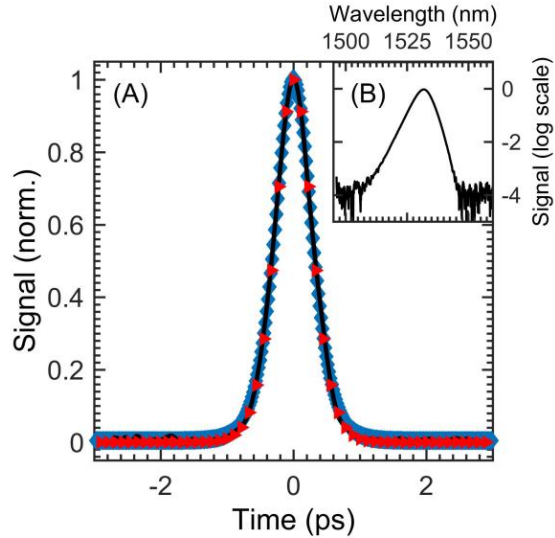

Fig. 3. Ultrafast pulse generation at 1.3 GHz repetition rate. (A) Experimentally (Black, solid line) observed autocorrelation trace, $f(t)$ fit (blue, diamond) to the AC trace and Fourier transform (red, triangle) of the mode-locked spectral power density. (B) Experimentally observed mode-locked spectral power density.

## 4. Conclusions

We demonstrate a passively mode-locked Er-Yb-ZBLAN waveguide laser oscillator using a semiconductor saturable absorber. This waveguide oscillator produces mode-locked pulses with 25 nm bandwidth utilizing the wide gain bandwidth of erbium. To the best of our knowledge, this is the first experimental evidence to support more than 10 nm spectral bandwidth (FWHM) in a passively mode-locked waveguide laser. With modest efficiency improvements, the cavity will be able to exploit the broader erbium bandwidth to achieve sub 100 fs pulses [37].

The performance of this mode-locked design should be directly transferrable to chip lasers that can be optimised to operate on a range of rare-earth transitions covering the near to short-wave infrared (1-3 µm) spectral domain. We are targeting peak-power scaling of SESAM mode-locked chip lasers to the multi kW regime with sub-100-fs pulse durations for high repetition rate compact frequency comb applications and non-linear frequency conversion.

## Funding

We appreciate Maptek Australia and Australian Research Council for the project support under the Linkage project LP130101133.

## Acknowledgement

T.M. Monro acknowledges the support of an ARC Georgina Sweet Laureate Fellowship.